\documentclass[journal]{IEEEtran}

\usepackage{amsfonts}
\usepackage{graphicx}
\usepackage{amsthm}
\usepackage{array}
\usepackage{bm}

\usepackage{amsmath}
\usepackage{subfig}

\usepackage{hyperref}

\begin{document}

\title{Large Language Model-Driven Classroom Flipping: Empowering Student-Centric Peer Questioning with Flipped Interaction}
\author{Chee Wei Tan\\
\thanks{The material in this paper was presented in part at the First International Workshop on Teaching Performance Analysis of ComputerSystems (TeaPACS 2021) In conjunction with the IFIP Performance 2021 Conference.

C. W. Tan is with the School of Computer Science and Engineering, Nanyang Technological University, Singapore 639798 (email: cheewei.tan@ntu.edu.sg).

Copyright (c) 2022 IEEE. Personal use of this material is permitted. However, permission to use this material for any other purposes must be obtained from the IEEE by sending a request to pubs-permissions@ieee.org.
}
\authorblockA{Nanyang Technological University, Singapore}}
\maketitle

%\thispagestyle{plain}
%\pagestyle{plain}
% As a general rule, do not put math, special symbols or citations
% in the abstract or keywords.

% IEEE
\begin{abstract}
Reciprocal questioning is essential for effective teaching and learning, fostering active engagement and deeper understanding through collaborative interactions, especially in large classrooms. Can large language model (LLM), such as OpenAI's GPT (Generative Pre-trained Transformer) series, assist in this? This paper investigates a pedagogical approach of classroom flipping based on flipped interaction in LLMs. Flipped interaction involves using language models to prioritize generating questions instead of answers to prompts. We demonstrate how traditional classroom flipping techniques, including Peer Instruction and Just-in-Time Teaching (JiTT), can be enhanced through flipped interaction techniques, creating student-centric questions for hybrid teaching. In particular, we propose a workflow to integrate prompt engineering with clicker and JiTT quizzes by a {\it poll-prompt-quiz routine} and a {\it quiz-prompt-discuss routine} to empower students to self-regulate their learning capacity and enable teachers to swiftly personalize training pathways. We develop an LLM-driven chatbot software that digitizes various elements of classroom flipping and facilitates the assessment of students using these routines to deliver peer-generated questions. We have applied our LLM-driven chatbot software for teaching both undergraduate and graduate students from 2020 to 2022, effectively useful for bridging the gap between teachers and students in remote teaching during the COVID-19 pandemic years. In particular, LLM-driven classroom flipping can be particularly beneficial in large class settings to optimize teaching pace and enable engaging classroom experiences.
\end{abstract}

% Note that keywords are not normally used for peerreview papers.

% % For peer review papers, you can put extra information on the cover
% % page as needed:
% % \ifCLASSOPTIONpeerreview
% % \begin{center} \bfseries EDICS Category: 3-BBND \end{center}
% % \fi
% %
% % For peerreview papers, this IEEEtran command inserts a page break and
% % creates the second title. It will be ignored for other modes.
% \IEEEpeerreviewmaketitle

%\input{introduction}
%\input{model0}
%\input{model1}
%\input{model2}
%\input{application}
% \input{experiment}
%\input{userstudy}
%\input{conclusion}

\section{Introduction}
\label{sec:intro}

In recent times, universities have taken a prominent role in adopting cutting-edge technologies to enhance the quality of teaching and learning \cite{wieman2017improving,deslauriers2011improved}. The integration of technology into educational practices has become evident through the widespread use of computer-assisted assessments, Massive Open Online Courses (MOOCs) like Coursera, and remote video-conferencing tools such as Zoom. Lecture time is less essential to content delivery than it once was, as interaction between the teacher and students becomes more important in the classroom. Specifically, classroom flipping in blended learning proves advantageous for students.

Classroom flipping is characterized by its emphasis on bridging the gap between teachers and students by fostering collaboration among peers. Nonetheless, the COVID-19 pandemic intensified this gap as higher learning institutions worldwide were forced to rapidly shift to online teaching and continue to maintain certain aspects of remote instruction even after the pandemic subsided \cite{lancet,nature,mitchell}. This abrupt transformation has posed an unparalleled challenge for universities globally as they grapple with reconciling the disparities between traditional in-person teaching and the new paradigm of blended learning. In addition to the challenges posed by the COVID-19 pandemic, the landscape of education witnessed a significant breakthrough in 2020 with the arrival of Large Language Models (LLMs) like OpenAI's GPT (Generative Pre-trained Transformer) series \cite{brown2020language,openai_2023}.\footnote{During the COVID-19 pandemic years, from 2020 to 2022, the advancement of Large Language Models (LLMs) marked two significant milestones. On June 3, 2020, GPT-3, equipped with an impressive 175 billion parameters, showcased remarkable language understanding capabilities \cite{brown2020language}. Subsequently, on November 30, 2022, ChatGPT (Chat Generative Pre-Trained Transformer), a breakthrough model built upon GPT-3.5 and GPT-4, emerged and revolutionized the field by excelling in diverse tasks without extensive fine-tuning \cite{openai_2023}.} These cutting-edge language models, exemplified by GPT-3 and others, have revolutionized the field of natural language processing (NLP) and artificial intelligence (AI). Advancements in reinforcement learning-based algorithms and access to vast datasets have been pivotal in developing LLMs \cite{christiano2017deep,vaswani2017attention}. Notable models like OpenAI's GPT series \cite{openai_2023}, Google's BERT Transformers \cite{bard_2023} exhibit exceptional accuracy in comprehending complex language patterns and generating human-like responses using prompt engineering techniques \cite{brown2020language,ouyang2022training}. 

In this context, the role of AI technologies in facilitating classroom teaching becomes increasingly critical \cite{dreyfus}. Educators seeking to automate teaching processes should be interested in a comprehensive theory of questions \cite{erotetics0,erotetics1}. A well-developed theory would bring us closer to the ideal of programming machines to handle all questions, respond appropriately, and assess student responses. Additionally, such machines could generate new and engaging assessment questions efficiently \cite{erotetics2}. This paper delves into the importance of generative AI technology, specifically Large Language Models (LLMs), in facilitating successful classroom flipping and closing the gap between instructors and students. It emphasizes the use of prompts, which are instructions to LLMs that elicit responses in the correct and acceptable form. We focus on prompt engineering to generate student questions and explores innovative approaches for seamlessly integrating LLM technologies with classroom flipping pedagogical methods.

Fostering an environment that encourages peer questioning can mitigate potential negative experiences for students and maintain the value of online learning to some extent. An effective classroom flipping approach thus must encourage some form of peer collaboration. For example, students first watch online lectures and then engage in peer discussions guided by questionings. Another example is Peer Instructions whereby students work individually and also with peers to address clicker questioning. In all these cases, teachers have to design well-conceived questions and feedback to assess the efficacy of learning \cite{feedback,tan2021}. Given the scarcity of real-time data, it is imperative to ensure that classroom flipping, especially with LLM integration, preserves the quality of teaching and learning. This raises the question: Can LLM techniques innovate peer questioning in classroom flipping, particularly for large classes?

In this paper, we explore how LLM techniques facilitate reciprocal peer-questioning for collaborative learning that can be seamlessly integrated with traditional evidence-based pedagogical methods like Peer Instruction and Just-In-Time Teaching. Based on student response data collection (.e.g, response to polls and quizzes) on their personal mobile devices, we utilize prompt engineering and LLMs to generate questions that can bridge both synchronous and asynchronous learning activities, providing a streamlined learning experience.  Another contribution of this paper is a LLM-driven mobile chatbot technology to accompany this LLM-based classroom flipping pedagogy, innovating content delivery in the form of ``low-stakes'' assessment, provide fast real-time feedback to students and for teachers to read their classroom better. It is expected that LLM-driven mobile chatbot technology can close the loop between knowledge delivery and its assimilation. We also discuss some open issues and potential challenges arising from this LLM-driven approach to pedagogy.

This paper is organized as follows. We first give an overview of various forms of classroom flipping in Section \ref{sec:classroomflipping}. Then, we discuss how large language models and flipped interaction techniques can be used together with classroom flipping pedagogy like peer instruction and just-in-time teaching in Section \ref{sec:llm}. We introduce our mobile chatbot software that integrates with these classroom flipping approaches in Section \ref{sec:chatbot}. Lastly, we summarize several key issues and open problems in Section \ref{sec:openissues} and we conclude the paper. The software and data for the experiments in this paper are available at: \href{https://github.com/henryherrington/nemobot-labs}{https://github.com/henryherrington/nemobot-labs}.

\section{Classroom Flipping by Reciprocal Questioning}
\label{sec:classroomflipping}

\begin{figure*}[ht]
\centering
\includegraphics[width=\linewidth]{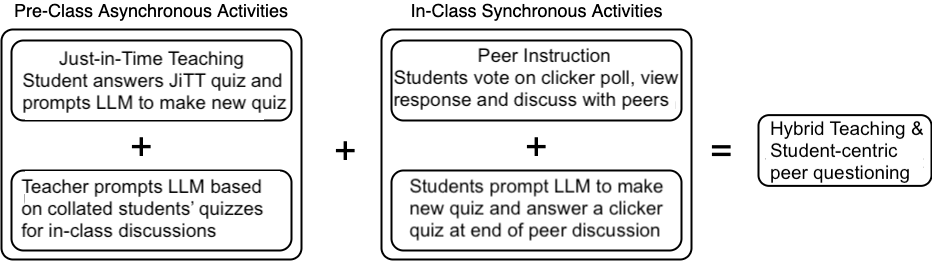}  
   \caption{An overview of classroom flipping based on large language models where a teacher blends Just-in-Time Teaching and Peer Instruction pedagogy together. This hybrid teaching fusion enables the inclusion of student-centered peer questioning through the utilization of prompt engineering in language models, allowing teachers to control the teaching pace effectively. }
   \label{fig:structure}
\end{figure*}

Classroom flipping is a pedagogical approach that emphasizes engaging students in active learning during class time rather than solely relying on traditional lectures. Instructors design short learning activities lasting five to ten minutes, reducing lecture time. To supplement content delivery, students are assigned pre-class activities like watching videos, reading assignments, and quizzes, while the class itself focuses on group activities enriched with interactive elements. A fundamental aspect of classroom flipping revolves around feedback through questioning, encompassing interactions where the instructor poses questions to students and students engage in peer questioning. Various approaches exist, including Peer Instruction \cite{mazur,lasry2008peer,deslauriers2011improved,wieman2017improving}, Just-In-Time teaching \cite{jit} and hybrid teaching \cite{watkins2010jittpi}. In this section, we show how reciprocal peer-questioning, a student-centric learning strategy where students create questions about the material they are learning \cite{reciprocalq1,reciprocalq2}, can be integrated with classroom flipping using LLMs for hybrid teaching as shown in Figure \ref{fig:structure}.

\subsection{Reciprocal Peer Questioning}

In a guided reciprocal peer-questioning approach, students are given a set of generic questions to help them generate their own specific questions. They then work in small groups to ask and answer each other's questions. From a metacognitive perspective, it is found that students who used guided reciprocal peer-questioning asked more critical thinking questions, provided more elaborate explanations and demonstrated higher better learning outcomes than students who used other approaches \cite{reciprocalq1,reciprocalq2}. The role of the instructor is thus to control the quality of questioning, ensuring that students are asking questions that are meaningful and thought-provoking. We discuss in the following how guided reciprocal peer-questioning is used in the context of traditional classroom flipping approaches and can foster peer interaction. We then discuss how LLMs can play a role to assist jointly the students and the instructor to controlling the quality of questioning in the next section, thereby influencing peer responses positively. 

\subsection{Peer Instruction: Poll, Prompt, Quiz}

Peer Instruction, as pioneered by Eric Mazur in \cite{mazur,lasry2008peer} and Carl Wieman \cite{deslauriers2011improved,wieman2017improving}, encourages students to share their understandings with peers who in turn challenge their interpretation during group activities. Teaching with peer instruction relies on the use of clickers -– a kind of audience response system -- to query students in classes and encourage them to seek out peers with different perspectives on a question to discuss before giving them the correct solution. Students are initially presented with a question and asked to vote individually. Subsequently, students can view the whole-class response displayed through statistical illustrations like histograms or pie charts, and engage in peer discussions, reflecting on their votes. Afterwards, a low-stakes quiz, mirroring the initial poll question, is conducted, followed by the instructor delivering a lecture on "foundational knowledge" pertaining to the poll-quiz process. This can be repeated a few times within a single lecture, and enables students to receive peer feedback while the instructor gathers feedback from the entire class, synchronously facilitated by clickers. 

Clicker polls and quizzes are often multiple-choice questions (MCQs) for ease of delivery. Peer discussions flourish when the questions are designed to have multiple defensible answers. Rather than picking a single correct option from several choices, students are asked to briefly discuss with their neighboring peers regarding the options and their poll responses. However, crafting a well-suited pair of poll and MCQ for peer instruction requires artful consideration \cite{taomultiplechoice,hardy2014student}. Patrick H. Winston, an AI expert noted for teaching excellence at MIT, explains clicker-enabled polling \cite{winston}:

{\it ``One obvious advantage is that clicker polling does not embarrass shy students fearful of ridicule if they choose the wrong answer. One not-so-obvious advantage is that instructors who choose to have clickers feel obligated to use them, and so must conceive interesting and informing polling questions."} \\
(Quote from Patrick H. Winston, 2020 \cite{winston})

Indeed, instructors will be motivated to find interesting and engaging questions to create the right combination of poll-quiz assessment that synchronizes with in-class teaching. We propose a {\it poll-prompt-quiz routine} to incorporate guided reciprocal peer-questioning with LLMs whereby, during the small-group peer discussion stage, students utilize the instructor-provided clicker quiz as a guide to generate their own task-specific questions (another MCQ related to the clicker poll) as outlined under the in-class synchronous activities in Figure \ref{fig:structure}. 
 Students are motivated to generate and answer their own LLM-generated questions. Instructors actively participate by guiding students in creating prompts for LLMs to produce clicker questions. Once the peer discussion concludes, the student-created questions and prompts can be submitted online and collected in a central system. This scalable approach allows the integration of ``student-centric" questions with real-time feedback during peer instruction, facilitating an interactive and engaging learning experience.

\subsection{Just in Time Teaching: Quiz, Prompt, Discuss}

Another effective classroom flipping approach is Just-In-Time Teaching (JiTT), suitable for both in-person and hybrid teaching. The JiTT approach involves teachers adapting instruction based on feedback received before the class, catering to the interests and needs of the students \cite{jit}. By employing JiTT, instructors can bridge the gap with students and identify areas where students face challenges, allowing them to allocate class time more effectively. Typically, a brief diagnostic assessment, known as the JiTT quiz, is conducted before students come for the in-class teaching. Students can use their reading materials and online information to answer the JiTT quiz. Teachers then utilize the diagnostic results as discussion points during the class to actively engage students \cite{jit}. 

Besides feedback, the JiTT quiz can double as a low-stakes assessment when they are carefully designed. The responses to the JiTT quiz can be in open-ended format rather than requiring a specific correct or incorrect answer. The instructor can then obtain feedback by reviewing a summary of the JiTT quiz responses before the in-class teaching session. We propose a {\it quiz-prompt-discuss routine} in JiTT to incorporate guided reciprocal peer-questioning whereby students individually utilize the instructor-provided assignment (e.g., reading assignment plus short quizzes) as a guide to generate their own task-specific questions (JiTT quizzes) as outlined under the pre-class asynchronous activities in Figure \ref{fig:structure}. Students have to submit both the prompts and the LLM-generated questions, which are subsequently pooled centrally for the instructor use (as per the aforementioned poll-prompt-quiz routine). In addition, the instructor can utilize the response summary to query the LLM, thus obtaining a summarized outcome of students' expectations, aiding in class preparation. The instructor can consolidate responses from all past JiTT quizzes, enabling the instructor to query the LLM and receive automated recommendations to assist the instructor in optimizing teaching pace.

The Just-in-Time Teaching format can also be interpreted differently, presenting teaching materials ``bit by bit" with each ``bit" introduced precisely when related questions are raised \cite{chiangagile,chiang}. Rather than starting with technical knowledge, teachers may begin with engaging applications or talking points to motivate students. ``Foundational knowledge" is then taught as a natural outcome while answering questions that emerge from these structured discussions, fostering a conversational learning experience \cite{chiangagile,chiang}. This approach proves beneficial, especially for short online courses like MOOCs, and relies heavily on pre-class intelligence gathering such as the {\it quiz-prompt-discuss routine} to initiate fruitful classroom discussions. 

\begin{figure*}[htb]
\centering
\includegraphics[width=0.99\linewidth]{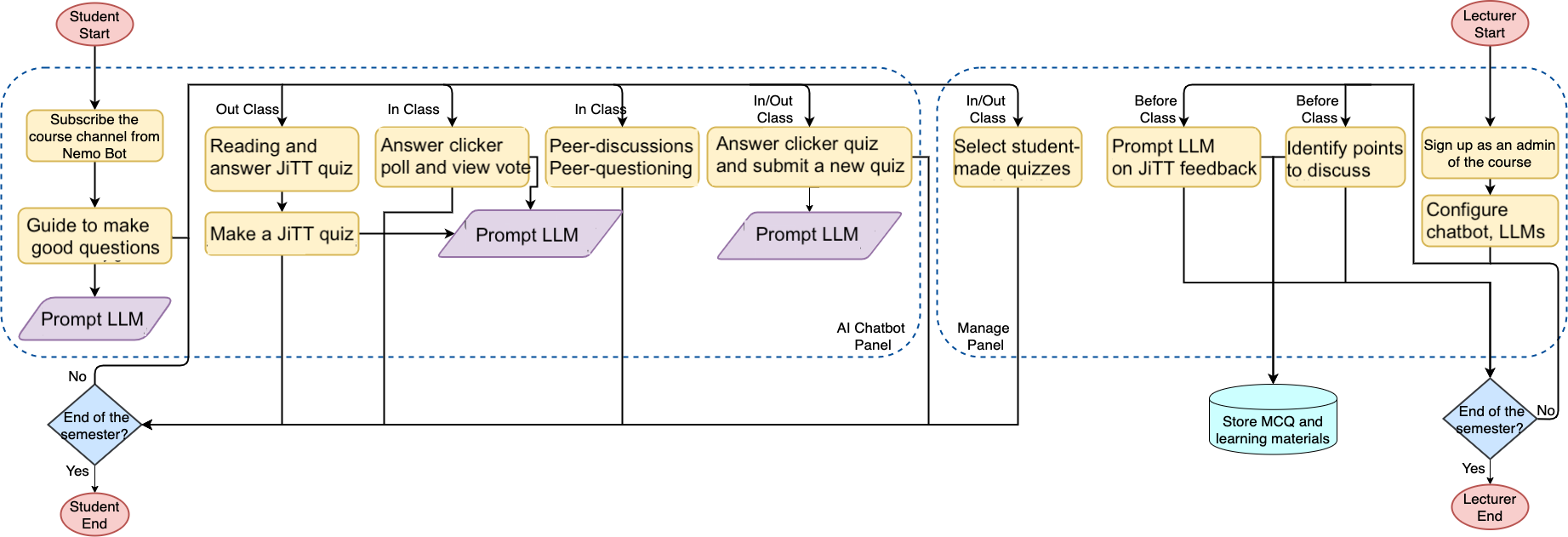}
\caption{Illustration of the workflow of the quiz-prompt-discuss routine and the poll-prompt-quiz routine to facilitate collaborative creation of multiple-choice questions by students through the flipped interaction of expansive language models and a mobile chatbot application (refer to Section \ref{sec:chatbot}) that can be employed by educators for classroom flipping.}
\label{fig:flowchart}
\end{figure*}

\subsection{Balancing Content Capacity and Comprehension Capacity}

We find that LLM-based techniques such as the ``poll-prompt-quiz routine" in peer instruction and the ``quiz-prompt-discuss routine" in JiTT can enhance communication between students and teachers in various classroom flipping approaches (see Section \ref{sec:chatbot} for details) as they enhance the real-time feedback between the instructor and students. Even in fully-remote classroom settings like MOOCs or during the COVID-19 pandemic, these routines can be adapted alongside automated audience response systems to offer instant feedback and remote guidance to students. Furthermore, {\it gamification} can be integrated to enhance these routines. Students can be 
 rewarded for identifying popular questions to be used in the next class, increasing engagement in remote teaching.

Indeed, motivating students to actively participate in reciprocal peer-questioning is essential at different learning stage of classroom flipping so as to encourage self-reflection and enable students to gauge their progress instantly.  Both the ``poll-prompt-quiz routine" and the ``quiz-prompt-discuss routine" empower students to self-regulate their learning capacity and enable teachers to swiftly personalize training pathways in real-time, unlocking the full potential of classroom flipping. Interestingly, the concepts behind these routines can be likened to the principles of AIMD/TCP in computer networking, enabling instructors to actively gather information about students' learning progress amid uncertainties \cite{tanaimd2021}. 

Let us draw a parallel between AIMD/TCP and flipped classroom teaching.\footnote{The author thanks Geoffrey M. Voelker for suggesting the parallel between AIMD/TCP and classroom teaching.} In the context of AIMD/TCP, users may desire to send as many data packets as possible without prior knowledge of the link capacity. In teaching, a similar concept can be referred to as ``{\it content capacity}," where instructors ask questions like ``{\it Am I teaching too much?}" or ``{\it Am I not teaching enough?}" Is there a need for a back-off when teachers realize they might be overwhelming students with content? Similarly, in AIMD/TCP, users are unaware of the actual link capacity and degree of fairness while sending data packets. In teaching, a corresponding term could be ``{\it comprehension capacity}," where teachers question whether some students are struggling and falling behind, or if the material is accessible to all students. Is there a slow start required when introducing new or advanced concepts in the classroom?  These are some of the key questions that are crucial to the Just-in-Time Teaching approach in \cite{chiangagile,chiang} that emphasizes delivering foundational knowledge contents in the nick of time rather than ``just in case," while balancing the tradeoffs between content capacity and comprehension capacity. The poll-prompt-quiz and quiz-prompt-discuss routines can be combined together to address these crucial questions to innovate flexible content delivery for hybrid teaching as shown in Figure \ref{fig:structure}.

\section{Flipped Interaction in Large Language Models}
\label{sec:llm}

We briefly introduce the Large Language Models. Among the most notable examples of LLMs are OpenAI's GPT series, including GPT-4 in ChatGPT \cite{openai_2023}, and other similar models like Google's BERT \cite{bard_2023}. The core foundation of LLMs lies in the transformer architecture, which was introduced in \cite{vaswani2017attention}. Transformers revolutionize the field of NLP through reinforcement learning-based algorithms and optimized transformer architectures that allow for efficient parallel processing, enabling significant scaling of the model size while maintaining computational tractability \cite{christiano2017deep}. The training process of LLMs usually involves a pre-training and fine-tuning phase. During pre-training, the model is exposed to vast amounts of diverse and unlabelled text data, which allows the learning of underlying structures and semantics of language. Once pre-training is complete, the model is fine-tuned on specific tasks using smaller, labeled datasets \cite{brown2020language,ouyang2022training}. Fine-tuning tailors the LLM to excel in particular applications, such as question-answering, sentiment analysis, language translation \cite{brown2020language} and even in AI-assisted programming \cite{wong2023natural}.

Despite the remarkable achievements of LLMs, they also come with certain challenges, particularly the AI alignment problem where hallucination in the context of AI refers to instances where the LLMs generate outputs that are not based on factual or accurate information but instead fabricate information \cite{aisafety1,aisafety2}. For example, an LLM may generate false statements or answer questions with made-up details that do not exist in reality. This issue is particularly concerning in language models because they have the potential to produce plausible-sounding but entirely fictional responses, leading to misinformation and miscommunication. Addressing hallucination is a critical aspect of AI safety and alignment to ensure that AI-generated content remains reliable and trustworthy \cite{aisafety0,aisafety1,aisafety2}. Ethical concerns, including biases present in the training data and potential misuse of these models for misinformation or malicious purposes, have also been subjects of intense scrutiny and debate \cite{aisafety0}. Hence, the incorporation of LLMs in classrooms necessitates instructors' oversight and guidance to ensure responsible usage by the students.

In the following, we delve into the specifics of LLMs for generating questions in the context of Peer Instructions and JiTT in Section \ref{sec:classroomflipping}. This involves finding suitable ways to prompt LLMs to generate questions, as illustrated in Figure \ref{fig:flowchart}. We also discuss advantages and limitations in using LLMs to facilitate collaborative reciprocal peer questioning. 

\subsection{Flipped Interaction Pattern}

The Flipped Interaction Pattern (FIP) technique represents a new approach that leverages the capabilities of LLMs to facilitate dynamic and interactive conversations with users \cite{flippedinteraction}. Traditional interactions with language models have typically followed a prompt-response paradigm, where users provide a query or statement, and the model generates a single, static response \cite{bengio2003neural}. The FIP technique is a flipped approach to achieve a specific goal. This is in contrast to the traditional way of using LLMs, whereby the user asks questions by specific prompts and the LLM provides answers. This inversion of the traditional interaction pattern creates a more interactive, back-and-forth exchange that can generate follow-up questions and responses during the interaction \cite{flippedinteraction}. Through the FIP technique, LLMs can elicit further clarification and explore additional dimensions of the user's inquiry based on the evolving context of the conversation. This shift from passive answering to active questioning imbues LLMs with the ability to engage users in more meaningful and insightful discussions, fostering deeper learning, and facilitating knowledge acquisition.

Effective prompts for the FIP technique adhere to two key principles: clarity and context. Clarity involves using simple and concise language that avoids complex vocabularies. Specific context is crucial for the LLM to understand the task and generate effective questions. Crafting prompts that lead to acceptable outcome requires careful consideration. The FIP technique can augment the model's contextual understanding by providing LLMs with more examples (known as few-shot prompting in machine learning terminology). This enlarges the potential for follow-up questions that users may desire and prevents the model from getting stuck in repeating the same question or repetitive low-diversity questioning cycles.

\subsection{Reciprocal Questioning For Students, by Students}
Empowering undergraduate students to formulate reciprocal peer questions using LLMs can be educationally enriching \cite{mcqgpt1}. Allowing students to take the driver's seat in creating clicker and JiTT quizzes using FIP encourages them to think critically and thoughtfully about the course material, thus reinforcing their own understanding of the subject matter. However, students may initially struggle with the right prompts to produce quizzes that strike the right balance for the whole class. In addition, evaluating student-LLM interactions is challenging, as instructors cannot possibly monitor the prompting process by individual students.

A student-centric approach will require the instructor to play a guiding role, offering support and feedback to help students craft meaningful and relevant questions. For example, instructors can assist by providing prompt examples and sharing well-crafted prompts from students. Incorporating student-generated questions into low-stakes assessments can incentivize active participation and knowledge synthesis. The instructor can choose the most insightful questions for further discussions, reinforcing the collaborative nature of the learning process. Students not only improve their comprehension but also develop a sense of ownership and responsibility for their learning progress. An example of a FIP prompt for a quiz on basic number theory can be:

{\it ``Please ask me questions to help me understand the fundamental theorem of arithmetic. Once you have enough information, create a clicker quiz with four choices about the greatest common denominator."} 

Therefore, educators need to carefully consider the optimal approach for integrating LLM-based prompts, transforming this into a structured workflow process with sequential instructions that students are required to adhere to, as depicted in Figure \ref{fig:flowchart}. Utilizing LLM prompts repeatedly can result in generating more pertinent responses tailored to the student's preferences. This iterative process can lead to the desired type of questions, which can then be employed in the future for either the {\it poll-prompt-quiz} or {\it quiz-prompt-discuss} routines. Prompting LLMs to generate appropriate and contextually relevant follow-up questions requires must strike a balance between inquisitiveness of an individual student and relevance to the whole class. Instructors should thus have oversight in this workflow to ensuring that student-generated questions align with the overall relevant teaching objectives and also adhere to ethical guidelines on data privacy and the use of generative AI \cite{aisafety1,aisafety2}. Table \ref{tableprompts} shows a few examples of how LLM prompting is used to generate MCQ quizzes in a course on digital logic. 

Designing MCQ has been well-studied in education \cite{mcq0}. MCQs can serve as valuable aids for self-assessment in courses, especially when it comes to foundational material such as fundamental rules of calculation and basic derivations \cite{taomultiplechoice}. The re-prompting technique in FIP can also be adapted from evidence-based pedagogical methods to designing MCQ questions. An example is Ullman's {\it root question}, as described in \cite{Ullman}. A root question is a multiple-choice question intentionally crafted with multiple correct answers and several incorrect options. When students encounter a root question, they are presented with a specific problem to solve, and their comprehension of the answer is evaluated through a series of MCQs. In case of an incorrect response, students are offered a hint in the form of a choice explanation and encouraged to try again till they have completely solved all MCQs correctly (thereby solving the given specific problem). In our experience, designing root questions is non-trivial and requires dedicated effort, preferably done outside class time as shown in Figure \ref{fig:flowchart}.

Erotetics, or the logic of questions and answers (see \cite{erotetics0,erotetics1,erotetics2}), employs skillful questioning to prompt critical thinking, challenge assumptions, and promote self-discovery. It guides individuals to refine their thoughts and reach deeper insights through a structured dialogue of probing inquiries. Erotetics is relevant to instructional science when teachers incorporate it into their teaching. By utilizing concepts from question logic, teachers can assess and guide their instructional methods and proposed teaching programs \cite{erotetics0,erotetics1,erotetics2}. Concerning `why' questions, we typically expect the explanation to be logically derived from the item being explained, but there is no effective test for such logical derivability. Consequently, there is no effective way to determine the suitability of an answer to a `why' question. However, we do have effective methods for recognizing proofs, so it might be beneficial to specify that answers to `why' questions should include not just the premise that explains but the entire proof leading to the explained conclusion.

In the context of Just-in-Time Teaching (JiTT), when engaging in reciprocal peer-questioning, students may be prompted to elucidate the reasons behind an incorrect choice. The goal is for them to not only offer the rationale for the error but also provide insights into the approach used to rectify the mistake and suggest a potential problem-solving methodology. This methodology aids students in mastering the art of formulating their own queries and honing their ability to respond to inquiries. Furthermore, the teacher can supply students with cue words that facilitate the construction of questions pertaining to a specific subject. Several examples of such cue words include:
\begin{itemize}
    \item How are … and … alike?
    \item What are the strengths and weaknesses of …?
    \item What would happen if …?
    \item What is the evidence to support …?
\end{itemize}

\begin{table*}[htbp]
    \centering
    \begin{tabular}{p{0.1\linewidth} | p{0.4\linewidth} | p{0.4\linewidth} }
      Question Type  &  Prompts  & GPT-3 response \\ \hline
                     &           &                \\
      Clicker Poll 1 & Create a clicker poll for the topic of basic Boolean logic with two choices  & What is the output of the Boolean expression: NOT (A AND B)?
      
    A) A AND B
    
    B) NOT A OR NOT B
    
    Step-by-step solution:
    Start with the given expression: NOT (A AND B)
    Apply De Morgan's theorem: NOT A OR NOT B
    This is the final simplified expression, which means the correct answer is option B: NOT A OR NOT B. \\
      Clicker Poll 2 & Create a clicker poll on Boolean logic with four options that has two correct answers  & Which of the following Boolean expressions are equivalent to A OR (NOT B)?

A) NOT A AND B

B) A AND NOT B

C) NOT A OR B

D) NOT A AND NOT B

(Note: The correct answers are B) A AND NOT B and C) NOT A OR B)  \\ \hline
                     &           &                \\
      Clicker Quiz 1 & Create a clicker quiz with four choices (with the last choice as none of the above) based on the poll: What is the output of the Boolean expression: NOT (A AND
B)?

A) A AND B

B) NOT A OR NOT B

& What is the output of the Boolean expression: NOT (A AND B)?

A) A AND B

B) NOT A OR NOT B

C) A OR B

D) None of the above

(Note: The correct answer is B) NOT A OR NOT B) \\ 
      Clicker Quiz 2 & Create a clicker quiz with four choices on the topic of de Morgan's theorem for three Boolean variables  & Which expression represents De Morgan's Theorem for three Boolean variables (A, B, and C)?

A) NOT (A OR B OR C) = NOT A AND NOT B AND NOT C

B) NOT (A AND B AND C) = NOT A OR NOT B OR NOT C

C) NOT (A AND B AND C) = NOT A AND NOT B AND NOT C

D) NOT (A OR B OR C) = NOT A OR NOT B OR NOT C

(Note: The correct answer is B) NOT (A AND B AND C) = NOT A OR NOT B OR NOT C) \\  \hline
                     &           &                \\
      JiTT Quiz 1 \cite{nahin2012logician} & With an AND gate and a True constant, which essential gate is still needed to form a universal set? Explain your viewpoint. & Factually incorrect response with hallucinations\\
                    &           &                \\
      JiTT Quiz 2  & On the table before you are three small boxes, labeled A, B , and C. Inside each box is a colored plastic chip. One chip is red, one is white, and one is blue. You do not know which chip is in which box. Then, you are told that of the next three statements, exactly one is true:

Box A contains the red chip
Box B does not contain the red chip
Box C does not contain the blue chip
You do not know which of the three statements is the true one. From all this, determine the color of the chip in each box. Explain your answer step by step. & Factually incorrect response with hallucinations\\
    \end{tabular}
    \caption{Examples of Prompts and GPT-3 Question Generation for Clicker Polls, Clicker Quizzes and JiTT Quizzes for teaching logic in an undergraduate class on artificial intelligence. An example of a clicker quiz generated based on a related clicker poll is shown. }
    \label{tableprompts}
\end{table*}

The instructor has a role to play in guiding students on using FIP prompts so that the LLM can handle ambiguity and uncertainty in order to drive the conversation. This allows the LLM to focus on achieving a particular objective, such as generating a short quiz or generating an open-ended question. By reversing the interaction flow, the LLM can effectively determine the format, number, and content of interactions. Finally, the teacher can request students to write a one-page summary of their interactions with the LLM before the class (as per the {\it quiz-prompt-discuss} routine). The teacher can collect all the responses and use them to generate additional questions through FIP. These questions can be used in class activities or as graded assessments, resulting in quicker and more accurate outcomes and potentially tapping into knowledge that students may not have initially possessed. With tailored prompts and instructor guidance, each student can influence question generation and steer the language model's comprehension towards a collective class goal.

To foster critical thinking, JiTT questions should be open-ended and avoid simple yes-or-no answers. To achieve this, FIP prompt engineering involves scaffolding the first LLM-generated questions to match students' current understanding levels. Students are encouraged to vary the types of questions they generate, enabling them to reflect on the material from different angles. Furthermore, students are tasked with submitting their own JiTT questions, including LLM-generated ones, along with references, empowering them to take ownership of their learning journey. Table \ref{tableprompts} contains examples of prompts and GPT-3 response to the generation of clicker polls, clicker quizzes and JiTT quizzes in an undergraduate class on artificial intelligence. Examples of clicker and JiTT quizzes generated using FIP along with incorrect LLM responses reported by students are given in Table \ref{tableprompts}.

\subsection{A Hypothetical Classroom Example}
We describe how to use the flipped interaction pattern in LLMs for reciprocal peer questioning together with the quiz-prompt-discuss and poll-prompt-quiz routines.

In teaching her large junior undergraduate class on introductory artificial intelligence, Professor Higgins aims to foster an engaging and intellectually stimulating learning environment. When covering the topic of Boolean algebra and logic, she draws inspiration from the work of Gardner \cite{gardner1992boolean}, focusing on connecting Boolean logic with fundamental concepts that first-year students might readily comprehend. Recognizing the potential of LLMs to enrich classroom discussions, she decides to integrate the {\it poll-prompt-quiz} routine as reading assignment and the {\it poll-prompt-quiz} routine for clicker teaching to enhance their understanding of Boolean logic.

{\it Step 1: Introduction to Classroom Flipping using LLMs}

Professor Higgins begins her first class by introducing LLMs and their roles in weekly assigned reading and clicker teaching using peer instructions. She explains how these powerful language models can understand basic concepts in mathematics and computer science, and generate human-like responses. Next, she introduces LLM prompting in the {\it poll-prompt-quiz} and the {\it poll-prompt-quiz} for outside-classroom and in-class activities respectively.  She highlights the standard expectations for students participating in these two routines. To motivate the students, she mentioned that LLM-based question submission will be assessed based on the effort put forth. 

{\it Step 2: Poll-Prompt Process during Peer Instruction}

In the peer instruction sessions, Professor Higgins conducts a clicker poll (e.g., Clicker Poll 1 in Table \ref{tableprompts}) and displays the combined responses after collecting individual polls. She then forms small peer discussion groups consisting of approximately two to three students each. Each group is tasked with creating a clicker quiz using reciprocal peer questioning. They generate quizzes that they can ask among themselves while Professor Higgins listens in to their conversations, moving around the groups.

{\it Step 3: Prompt-Quiz Process during Peer Instruction}

After the peer discussions, Professor Higgins allows each group to access a shared system for inputting their clicker questions (see Section \ref{sec:chatbot} later). Once this is done, she administers a clicker quiz (e.g., Clicker Quiz 1 in Table \ref{tableprompts}) to all students individually, where they submit their answers. Professor Higgins informs the class that future clicker quizzes may include questions generated by the students themselves.

{\it Step 4: Quiz-Prompt Process during JiTT Quiz}

Students are tasked with some reading assignments (Chapter 8 in \cite{gardner1992boolean} and Chapter 4 in \cite{nahin2012logician} on Boolean logic) and given a JiTT quiz (e.g., both JiTT Quizzes 1 and 2 in Table \ref{tableprompts}), and asked to submit online a LLM-generated question related to logic puzzles and a one-page interaction between the student and the LLM. With the collated responses, Professor Higgins queries the LLM on talking points for a discussion between Boolean algebra and basic number theory, and also applies FIP to generate follow-up questions that can be used for clicker teaching.

{\it Step 5: Prompt-Discuss Process during In-Class Discussions}
Professor Higgins and her students participate in dynamic discussions prompted by the talking points provided by the LLM in Step 4. During these discussions, she encourages some students to present their comprehension in front of the class. Occasionally, Professor Higgins surprises the students by revealing that the JiTT quiz they took was actually created and submitted by one of their fellow students. She also asks the student to share briefly some tips on LLM prompting to improve the quality of quiz-generation. For instance:

Student 1: ``{\it With an AND gate and a True constant, which essential gate is needed to form an OR gate?}"

Professor Higgins: ``Indeed, LLM returns a seemingly correct answer when it isn't? But can you see how it makes a different attempt if I rephrase the JiTT quiz as ``{\it Try constructing an OR gate using only an XOR and AND gate}"?

Student 2: ``The LLM is {\it hallucinating}! It produced an incorrect answer to my JiTT quiz (i.e., JiTT Quiz 2 in Table \ref{tableprompts}) despite giving an extensive and seemingly-correct argument." 

Professor Higgins: "Can you identify and correct the logical fallacy through additional prompts? See that the LLM can answer satisfactory a logic puzzle with {\it two} variables, yet it struggles when handling a puzzle with {\it three} variables."\footnote{From our observations, GPT-3 spectacularly failed on this puzzle in JiTT Quiz 2 in Table \ref{tableprompts} with endless hallucinations, while GPT-4's failures are less pronounced. In our view, logic puzzles serve as useful benchmarks for LLMs.}

{\it Step 6: Whole-Class Discussion and Reflection}

To conclude the activity, the class comes together for a whole-class discussion. Each group presents their key insights and highlights the most intriguing questions generated by the LLM during their discussions, and discuss how the LLM-driven questions prompted them to explore different aspects of their assigned topics and encouraged them to see the deeper connections between seemingly-unrelated concepts: the greatest common denominator and the least common multiple are conceptually similar to the Exclusive-OR (XOR) and the AND logic operations in Boolean logic \cite{gardner1992boolean}.  

Moreover, students also reflect on the potential risks of relying solely on LLM-generated questions and how it relates to AI ethics. They consider the scenario where they or Professor Higgins might overlook errors in the LLM-generated questions and accept them as valid quiz questions without scrutiny.  To address this, students readily agree with Professor Higgins's suggestion that they should be offered bonus points if they can identify any such errors in JiTT or clicker quizzes, encouraging them to be vigilant and critical in their learning process. 

{\it Epilogue}:
At the end of her course, Professor Higgins presented her students with an intriguing challenge: to discover the prompts, or a sequence of prompts, that would guide a LLM (e.g., GPT-4 in ChatGPT) to correctly solve, or at least approach the correct solution, to the puzzle presented in Shannon's paper \cite{shannon1953computers}: {\it It is known that salesmen always tell the truth and engineers always tell lies. G and E are salesmen. C states that D is an engineer. A declares that B affirms that C asserts that D says that E insists that F denies that G is a salesman. If A is an engineer, how many engineers are there?}\footnote{This logic puzzle given in \cite{shannon1953computers} has seven Boolean variables and requires advanced questioning using the process of elimination. Above all, it exudes the importance of the {\it law of excluded middle}: for every proposition, either this proposition or its negation is true. To tackle LLM's hallucinations, consider adopting negation as failure, where propositions are either true or unprovable, rather than strictly true or false, from our perspective.}

By incorporating LLMs into the reciprocal peer questioning process for Boolean logic, Professor Higgins empowers her students to engage dynamically with numerous computer science concepts and mathematics. The author adopted this workflow as summarized in Figure \ref{fig:flowchart} for teaching introductory artificial intelligence courses for computer science freshmen and postgraduate courses on convex optimization from 2020 to 2022. These university courses were newly-offered in 2020, coinciding with the COVID-19 pandemic when online instruction became the standard practice. The author found that the use of LLMs can stimulate critical thinking as students did raise more questions on the validity of LLM-generated questions, leading to more insightful discussions during class time. The reciprocal peer questioning activities thus encouraged collaborative peer questioning, enabling students to benefit from their peers' perspectives. Lastly, the practical exposure to LLM-driven chatbot technology (described in details in the next section) can prepare students for the evolving conversational AI landscape in the age of ChatGPT.

\section{Mobile Chatbot Software Implementation}
\label{sec:chatbot}

In this section, we discuss the implementation of a mobile chatbot software application in Facebook Messenger and Telegram, which are popular social messaging platforms. The contribution of this mobile chatbot software is two-fold: First, it digitizes the clicker in Peer Instruction and serves as a pacemaker for JiTT teaching and to blend all that together for hybrid teaching (cf. \cite{watkins2010jittpi}) as shown in Figure \ref{fig:structure}. Second, it enables the poll-prompt-quiz and poll-prompt-quiz routines to be implemented in hybrid teaching. The mobile chatbot software works with a central server backend that stores all the quizzes created by students who use a web interface as depicted in Figure \ref{fig:quizpool1}(a), while the instructor and teaching assistants use a web interface to vet these questions as depicted in Figure \ref{fig:quizpool1}(b), which also acts as the dashboard for the instructor to push out the clicker poll and quizzes.

\subsection{Digitizing the Peer Instruction Clicker}

Clickers are commonly used feedback tools for peer instruction in numerous higher learning institutions \cite{martyn2007clickers}. The conventional clicker is a physical input device that employs a short-range wireless signal to establish a connection with a portable wireless receiver attached to the instructor's computer. Its primary function is to collect real-time responses from students, possibly without identifying individuals, enabling the instructor to share the response outcomes. Nevertheless, the traditional clicker has some drawbacks, such as costs, battery concerns, and the inability to be transferred between different classes or schools \cite{bruff2009teaching}. 

\begin{figure}
\centering
\subfloat[]{\includegraphics[width=0.45\columnwidth]{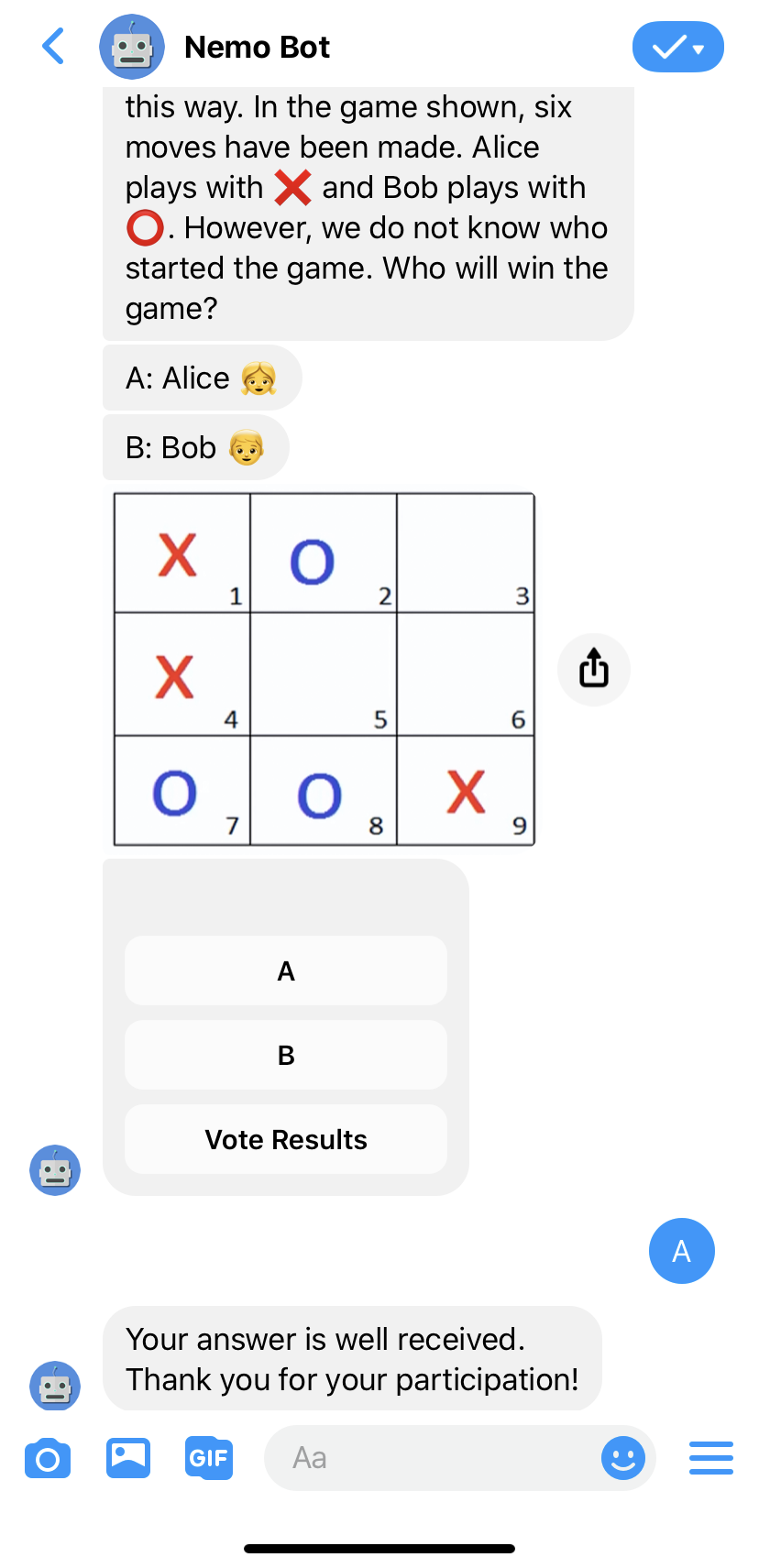}}
\hspace{1em}
\subfloat[]{\includegraphics[width=0.45\columnwidth]{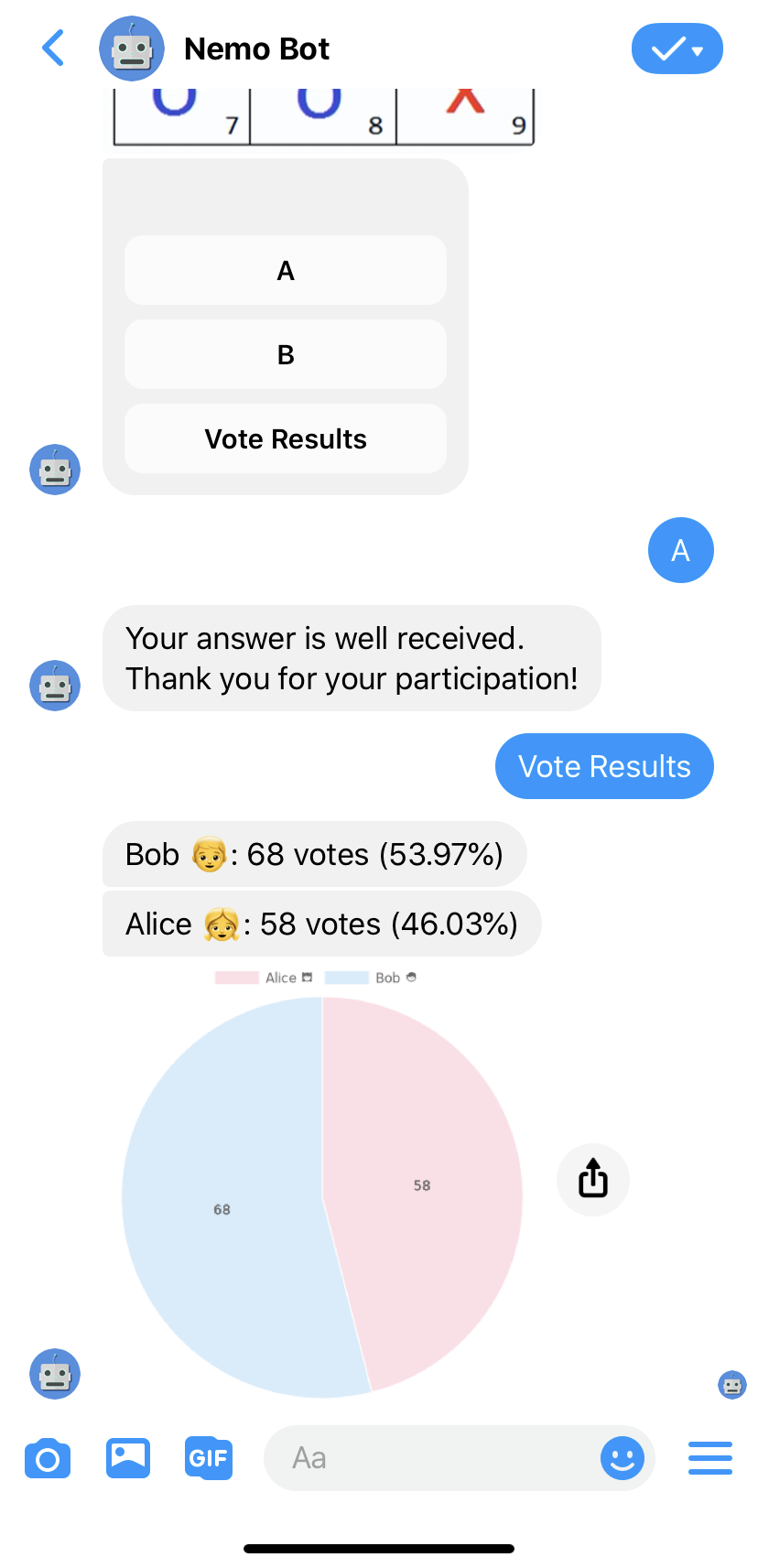}}
\caption{Illustration of the poll-prompt-quiz procedure utilizing mobile chatbot software that operates as a digital clicker within the context of Peer Instruction. The left side displays the poll quiz with student submissions, while the right side reveals the poll response following the student's vote submission. Subsequently, the student collaborates with peers to generate a comparable clicker poll using LLMs.}
\label{fig:clicker}
\end{figure}

The mobile chatbot software utilizes widely-used chat platforms like Facebook Messenger and Telegram, known for their user-friendly human-computer interface design. Within this software, all clicker polls and quizzes consist of MCQs. Students can instantly view the poll response outcome once they submit their vote. Since the clicker quiz question is linked to the poll question that will be discussed with peers, the clicker quiz thus serves as a low-stakes and relatively stress-free assessment. For the quiz assessment in the poll-prompt-quiz routine, a time-limit (e.g., five minutes) can be set for students to respond. An example of the human-computer interface is given in Figure \ref{fig:clicker}.

%\begin{figure}
%\centering
%\includegraphics[width=0.6\columnwidth]{figures/clicker3.png}
%\caption{Illustration of the whole-class response as seen by a student user who has participated in the poll with the mobile chatbot software functioning as a clicker in Peer Instruction.}
%\label{fig:clicker}
%\end{figure}

\subsection{Setting the Pace in JiTT}

\begin{figure}
  \centering
  % include first image
  \includegraphics[width=\linewidth]{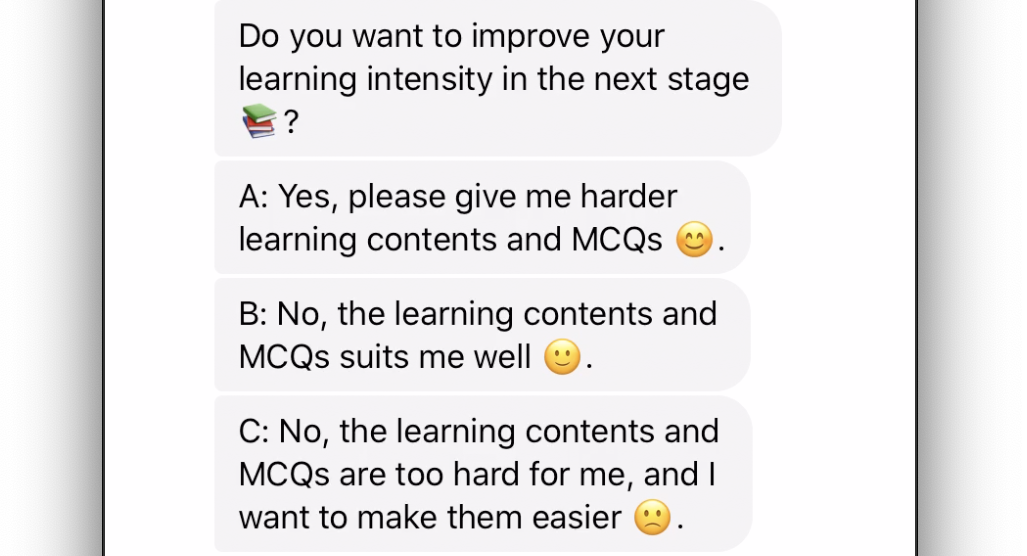}  
  \caption{Example of a message sent automatically by the chatbot to allow the student user to self-calibrate the difficulty level of the received JiTT quizzes. After completing the JiTT quiz, students are required to utilize LLMs to try creating a comparable quiz that they will submit.}
  \label{fig:dif}
\end{figure}

Similar to clicker quizzes, JiTT utilizes quiz diagnostics to determine the pace for both the teacher and individual students. The pre-class feedback query is often administered a couple of days before the actual teaching. When JiTT quizzes are employed as low-stakes assessments, instructors have the opportunity to design either MCQs or open-ended questions. In practice, students typically have more time and available resources for crafting JiTT quizzes beyond the confines of the classroom, in contrast to the immediate creation of clicker quizzes during class sessions. Our chatbot system autonomously issues regular prompts to students, prompting them to decide whether they want to engage with questions of moderate or elevated levels of difficulty as shown in Figure \ref{fig:dif}. 

\begin{figure}
\centering
\subfloat[]
{\includegraphics[width=0.45\columnwidth]{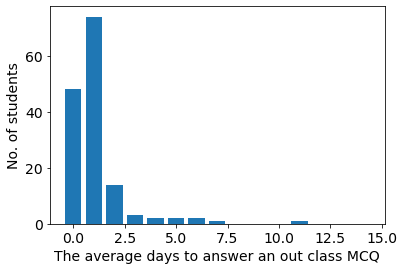}}
\hspace{1em}
\subfloat[]
{\includegraphics[width=0.45\columnwidth]{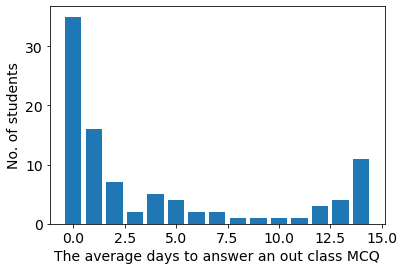}}
\caption{Average number of days to answering a student-designed JiTT quiz question.}
\label{fig:fig2}
\end{figure}

\begin{figure}
\centering
\subfloat[]{\includegraphics[width=0.45\columnwidth]{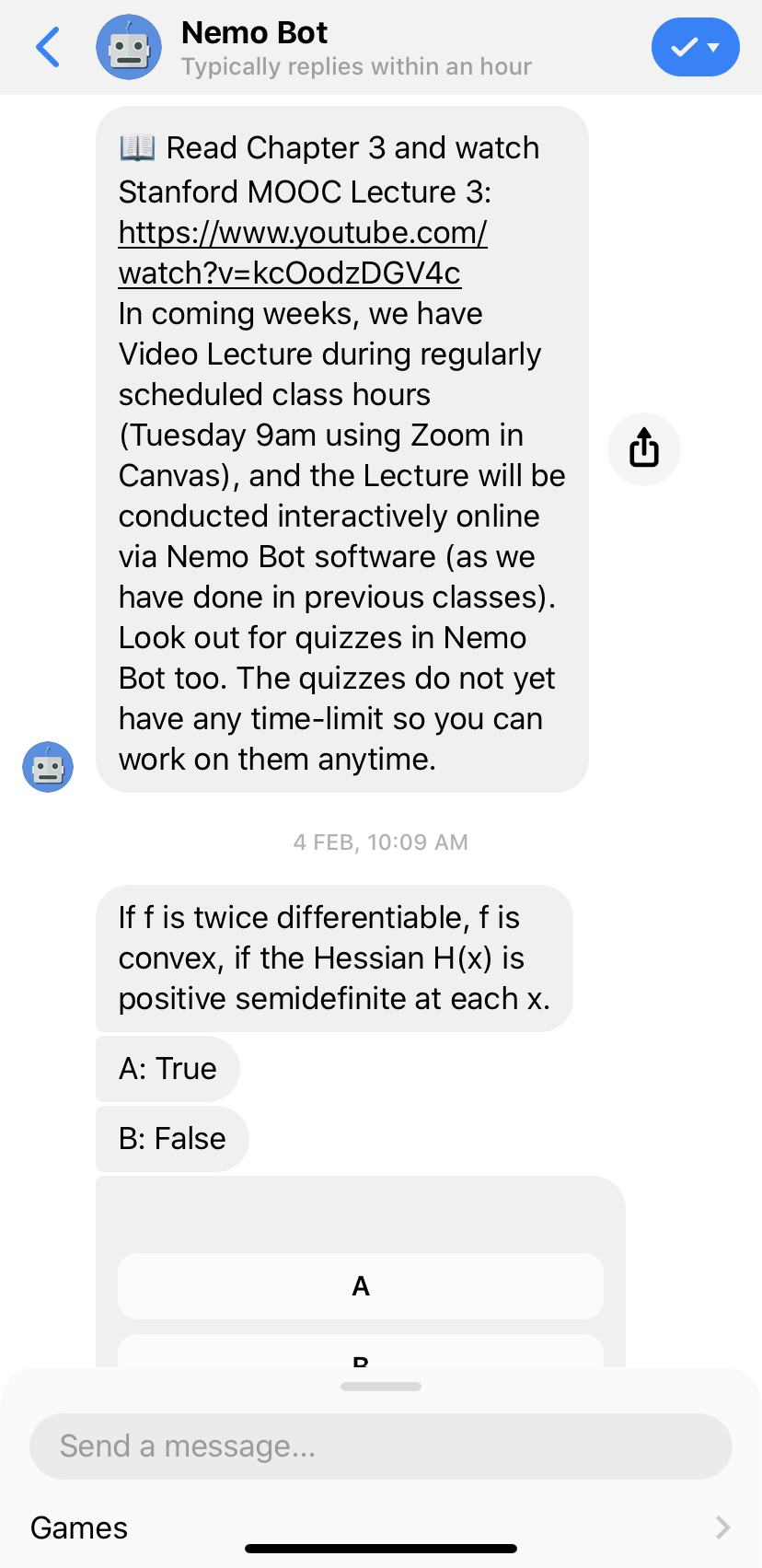}}
\hspace{1em}
\subfloat[]{\includegraphics[width=0.45\columnwidth]{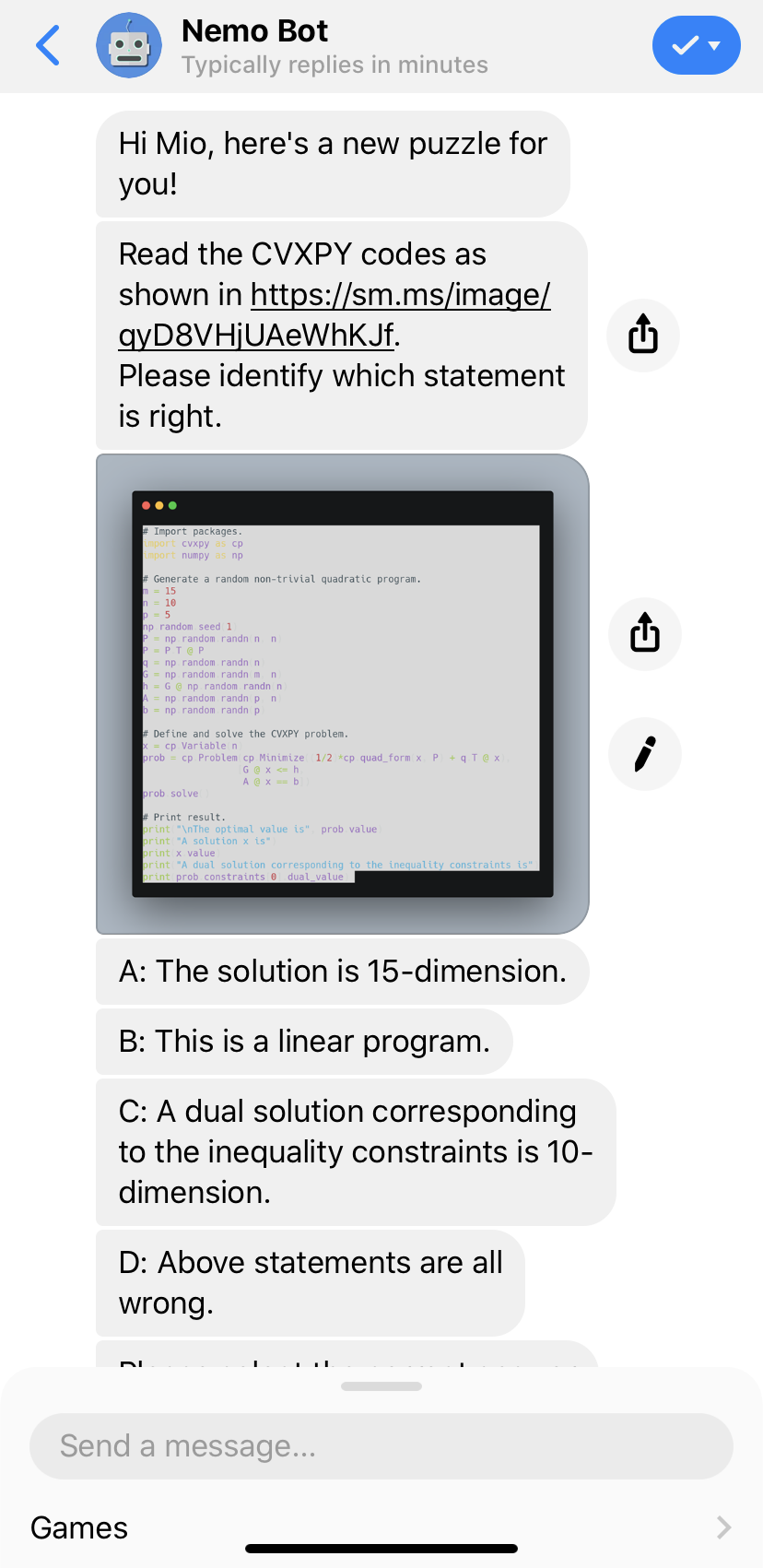}}
\caption{Illustration of a JiTT quiz in a quiz-prompt-discuss routine. A JiTT quiz does not impose any time restrictions on students for submitting their answers and grants them the flexibility to adjust the quiz's difficulty level based on their preference. In addition, the Codex LLM can be prompted using FIP to create a code snippet to accompany the JiTT quiz created by a student.}
\label{fig:jittquiz}
\end{figure}

Through monitoring the choices made by students to tackle more challenging questions, the teacher can extract valuable insights and feedback from the system. There is no time limit for JiTT quizzes in order to accommodate students residing at different time zones during the period of online teaching due to COVID-19. The chatbot is adaptable for both the poll-quiz routine and JiTT quiz (Figures \ref{fig:clicker} and \ref{fig:jittquiz}) by allowing MCQs with or without time limits to be configured. It is possible to integrate JiTT quiz creation with leaderboards and rankings by combining student response data to difficulty level of the JiTT quizzes as shown in Figure \ref{fig:dif}. This feature serves as an indicator to students regarding their peers' preferences for generating quizzes of specific difficulty levels for JiTT quizzes. It also enables students to seek assistance from their peers when tackling more demanding JiTT quizzes. We illustrate the duration in which a JiTT quiz gets answered by the students of an undergraduate class and a postgraduate class in Figure \ref{fig:fig2}(a) and (b) respectively. Figure \ref{fig:fig2} shows that most students are able to complete their submission in the first few days within a week from the day that the JiTT quiz is sent out.

\subsection{Integration of Chatbot with GPT and Quality Control}

The integration of the above chatbot interface is created using DialogFlow, which is Google's Conversational AI framework, that utilizes few-shot Natural Language Processing (NLP) models to discern the intent of conversational input \cite{dialogflow}. To set up intent understanding within Dialogflow, we establish an agent in our chatbot that involves defining user intentions (i.e., intents), supplying sample phrases, specifying keywords for data extraction, configuring responses, and conducting thorough testing to ensure accuracy and contextual comprehension \cite{dialogflow}. We first train these models on extensive datasets of text and code using DialogFlow to comprehend user queries, and we find that DialogFlow's few-shot NLP models offer a convenient tool to create usable chatbots for peer instruction and JiTT quiz administration with minimal training examples. However, the development and maintenance of conversational agents in Dialogflow still rely on pre-defined rules and handcrafted responses that can lead to limited interactions and rigid user experiences. 

\begin{figure}
\centering
\subfloat[]{\includegraphics[width=0.45\columnwidth]{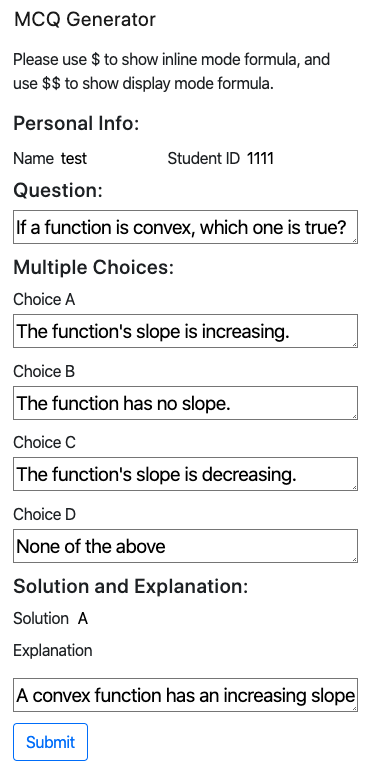}}
\hspace{1em}
\subfloat[]{\includegraphics[width=0.45\columnwidth]{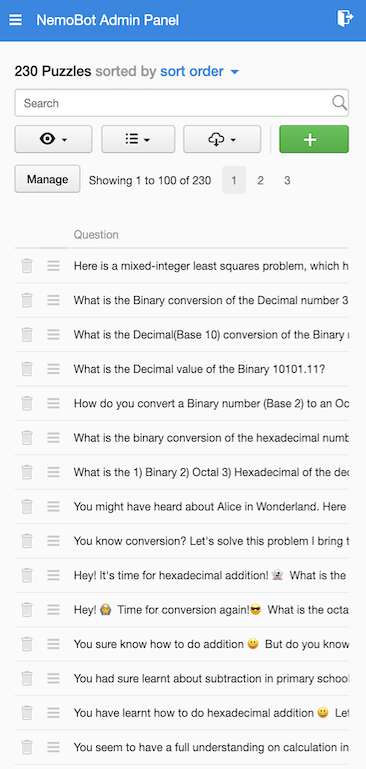}}
\caption{The dashboard panel that allows students to set their own poll or quiz for submission is shown on the left. On the right is the instructor dashboard panel for vetting all the submitted polls or multiple choice questions. The instructor can choose from these questions to be sent to the chatbot of all students.}
\label{fig:quizpool1}
\end{figure}

The chatbot integration with OpenAI's Language Models occurred in a few phases. By November 2021 when OpenAI made GPT-3 publicly-available, we asked the students to utilize the {\it OpenAI Playground} in \cite{openai_2023} for creating JiTT questions (and before that we use the GPT-3 API in beta version). To ensure the quality control in using GPT for FIP, we ask the teaching assistants to solicit the submission of GPT-generated questions and to reproduce the prompts in {\it OpenAI Playground}. If the GPT-generated questions are similar to that provided by the student, we store them into a central quiz database (as shown in Figure \ref{fig:quizpool1}) and configure the rules and intents in Google Dialogflow conversational agents accordingly. Thereafter, the 
 instructor can select and push out the quiz to the students in real time. 

Initially, we ask students to create clicker quizzes during the poll-prompt-quiz routine by assigning them to Zoom virtual classrooms to collaborate on GPT-generated quizzes using FIP and supervised by teaching assistants who have access to the GPT-3 API beta. Inspired by the virtual interaction, the students then work individually to submit their quizzes for vetting and storing in the central database. By leveraging the collective effort of teaching assistants, this approach can address limitations in using LLMs by some students and create a continual  peer learning experience during remote teaching. The {\it OpenAI Playground} thus allows the instructor and teaching assistants to continuously create more questions with the students' initial submission to bootstrap and enrich the quiz database based on students' evolving preferences gathered via the Google Dialogflow conversational agent.

Before the release of the OpenAI Playground, the OpenAI Codex API was released earlier in August 2021 \cite{chen2021evaluating}. Derived from GPT-3, OpenAI Codex's training corpus encompasses natural language and billions of lines of publicly-sourced code, including content from 54 million GitHub repositories \cite{chen2021evaluating}. The Codex exhibits proficiency in more than a dozen languages including Python, JavaScript, C and also other domain-specific languages, and is used in AI-assisted programming applications like the Copilot \cite{tan2023copilot,wong2023natural}. Though Codex has been depreciated since March 2023 and replaced by more advanced versions of the GPT (GPT-3.5 Turbo), the author expects the AI-assisted programming tool to be useful to educational domain specific languages besides general software development tasks like code translation and code generation \cite{tan2023copilot,wong2023natural}. In particular, we leveraged its code memory in the CVX code in Github repositories, enabling Codex-generated code snippets to be incorporated into the question generation for teaching our courses on convex optimization.\footnote{CVX in \cite{cvx,cvxpy} is an open source software developed by Stephen Boyd and his team at Stanford University for solving mathematical optimization problems by expressing them in domain-specific languages written in MATLAB, Python, or Julia to model convex functions and constraint sets \cite{boyd2004convex}.} 

\begin{figure}
  \centering
  % include first image
  \includegraphics[width=\linewidth]{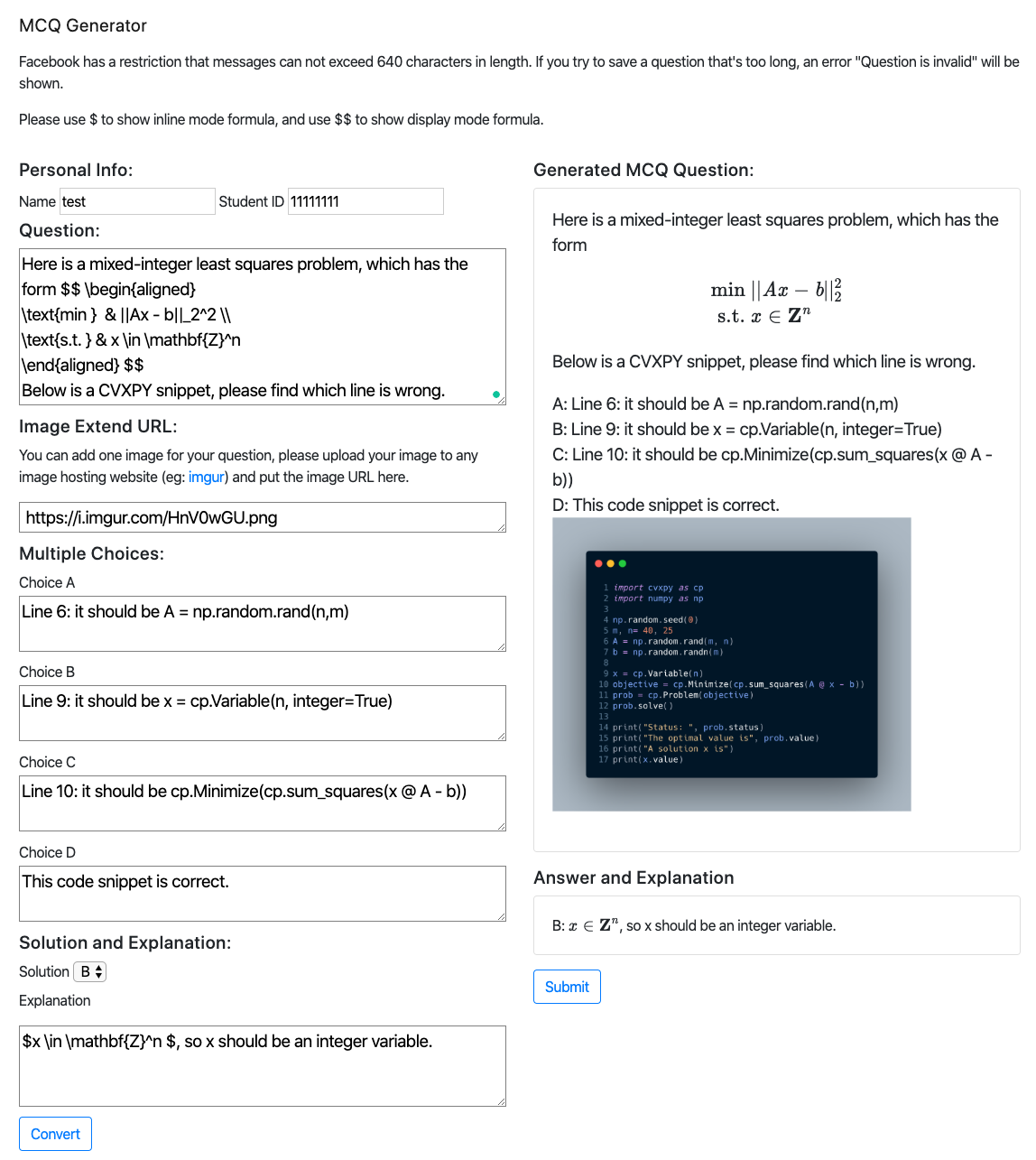}  
  \caption{Illustration of a JiTT quiz submission by a student who formulates a mathematical optimization problem question in Latex using OpenAI Playground (GPT-3) and its accompanying code snippet generated by the Codex for a class on convex optimization in 2022.}
  \label{fig:quizpool2}
\end{figure}

In our mobile chatbot software, these new MCQs can first be vetted quickly by the instructors via the instructor dashboard (or pre-filtered automatically in the backend) before they are used in the classroom. As an instance, to ensure the accuracy and reliability of integrating questions by students who use OpenAI Playground to generate a question in Latex,\footnote{LaTeX is a specialized domain-specific language widely used for typesetting and formatting documents, especially useful for handling mathematical equations that GPT-3 is capable of handling.} we ask the teaching assistants to prompt the Codex to generate the relevant CVX code to accompany the quiz, along with validation with the actual code output as shown in Figure \ref{fig:quizpool2}. The combined quiz then undergoes evaluation by the instructor who can use the chatbot system to get community-driven feedback during the prompt-quiz-discuss routine to help identify and rectify any misleading or erroneous question part. 

\begin{figure}
\centering
\subfloat[]{\includegraphics[width=0.45\columnwidth]{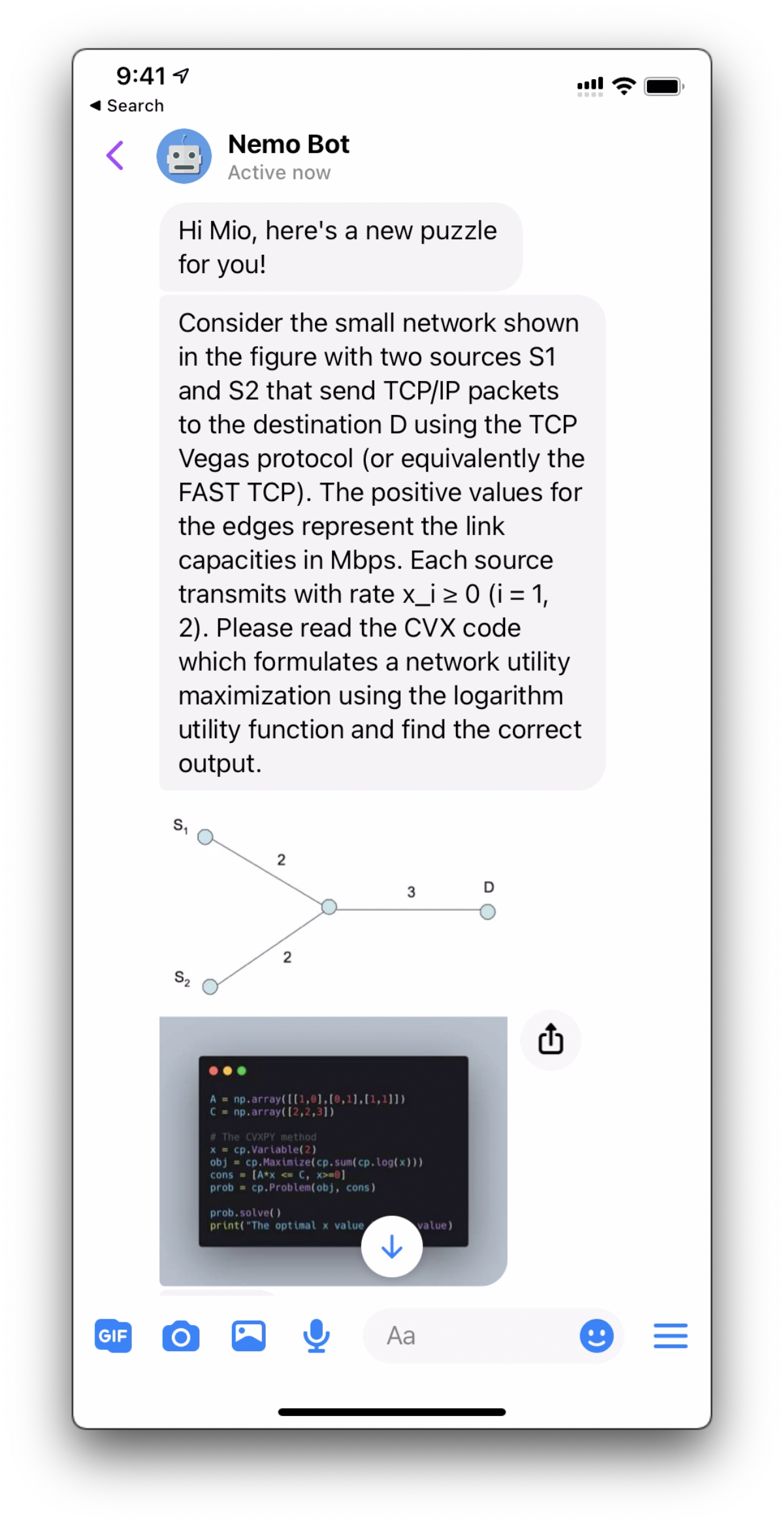}}
\hspace{1em}
\subfloat[]{\includegraphics[width=0.45\columnwidth]{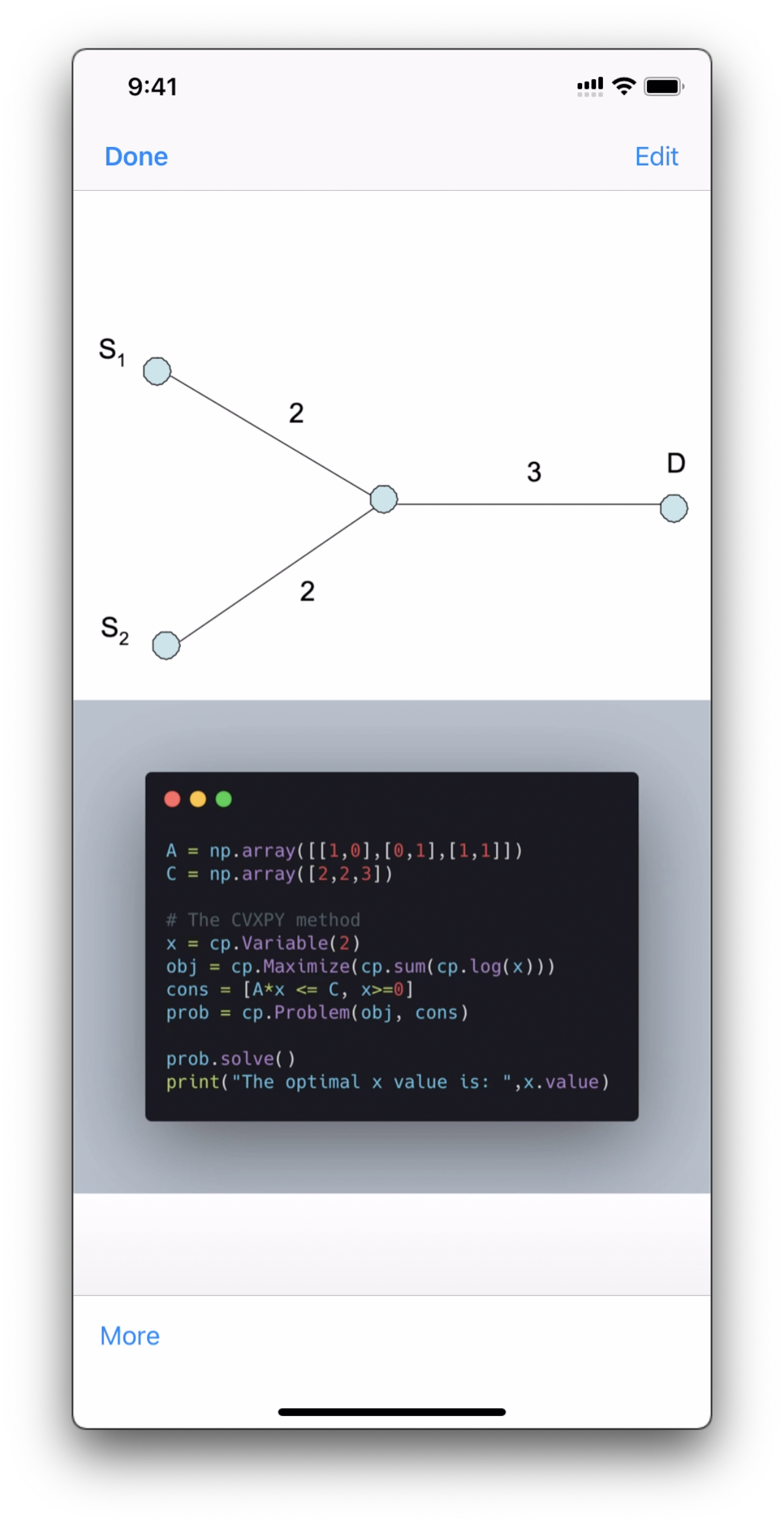}}
\caption{A JiTT quiz received via the mobile chatbot software on the topic of network utility maximization in (a) testing their understanding of a GPT-generated question on CVX \cite{cvx} with the accompanying code snippet generated by Codex and an image generated by a text-to-image generator, DALL-E, in (b).}
\label{fig:cvx2}
\end{figure}

This continuous feedback loop reinforces the reliability of reciprocal questioning using Codex in \cite{chen2021evaluating}, thereby mitigating potential bias constraints inherent in the two language models (i.e., Codex and OpenAI Playground) by ensuring human oversight over the quality of creating a clicker or JiTT quiz with accompanying code snippets. Furthermore, within the classroom's discussion phase, instructors can assess quiz quality and complexity, applicable to both clicker and JiTT quizzes in hybrid teaching. During this discussion phase, instructors can first assign preliminary difficulty scores to LLM-generated quizzes, and then later incorporate collective student performance to fine-tune the quiz difficulty scores (so that the quiz can be used for future cohorts). This adjustment process could involve collaborative filtering or crowd-sourcing algorithms, as described in \cite{autograding}. One can also integrate the OpenAI GPT-3 and Codex for quiz generation with improved presentation in our Dialogflow-based conversational AI chatbot framework. For example, one can use a text-to-image generation tool (e.g., OpenAI DALL-E) to generate informative images as shown in Figure \ref{fig:cvx2} to enhance students' learning experience during both the poll-prompt-quiz and quiz-prompt-discuss routines. Table \ref{tablesummary} illustrates the mean time taken by students to submit a GPT-generated quiz, along with the average difficulty level that they assign to their quiz on a scale of 1 to 10 (with 1 indicating easy and 10 indicating challenging). Based on our observations, students' end-of-course feedback reflects a favorable response to this unconventional hybrid teaching approach, particularly useful in remote teaching.

\begin{table*}[htbp]
    \centering
    \caption{Comparison of Hybrid Teaching Methods for 'Artificial Intelligence (GE2340)' Undergraduate Course and 'Convex Optimization and Its Applications' Postgraduate Course using LLMs from OpenAI between 2020-2022}
    \begin{tabular}{|c|c|c|c|c|}
        \hline
        \textbf{Courses} & \textbf{Artificial Intelligence (2020)} & 
        \textbf{Artificial Intelligence (2021)} & \textbf{Optimization (2021)}  &
    \textbf{Optimization (2022)}    \\
        \hline
        Class Type and Size & 
        182 Undergraduates & 
        167 Undergraduates & 
        63 Postgraduates & 
        57 Postgraduates \\
        \hline
        Teaching Format & 
        Online PI  & 
        Online PI and JiTT & 
        Online PI and JiTT & 
        Online PI and JiTT \\
        \hline
        LLM Model & 
        GPT-3 & 
        GPT-3 and DALL-E  & 
        GPT-3  & 
        GPT-3, Codex and DALL-E \\
        \hline
        Quiz Instances & 
        Figure \ref{fig:clicker}  & 
        Table \ref{tableprompts} & 
        Figure \ref{fig:jittquiz}  & 
        Figure \ref{fig:cvx2} \\
        \hline
        Time to Create Quiz & 
        2 days & 
        3 days  & 
        7 days  & 
        6 days \\
        \hline
        Difficulty Level of Quiz & 
        5 with variance 1.8  & 
        6 with variance 1.1 & 
        8 with variance 0.5 & 
        7 with variance 0.9 \\
        \hline
    \end{tabular}
    \label{tablesummary}
\end{table*}

\section{Open Issues}
\label{sec:openissues}
We highlight several open issues regarding the application of LLM technologies in classroom flipping, with a strong focus on advancing student learning excellence. Addressing these challenges will mark a significant stride in advancing LLM-based classroom flipping.

\begin{itemize}
    \item Questioning is vital in meaningful learning and scientific inquiry \cite{erotetics0}. One can explore how LLMs can embrace the logic of questions as discussed in \cite{erotetics1,erotetics2}. By doing so, LLMs can effectively integrate crucial pedagogical elements of classroom flipping to achieve the goal of {\it AI-assisted teaching}. It is also imperative to consider how bias concerns can be adequately addressed during this integration process.
    \item The study conducted by \cite{mitpress} underscored that students who faced difficulties with online teaching may be deterred from taking further courses in the same subject. Combining limited learner data, such as LLM prompts, even within a single course setting, can be intricate \cite{reich}. How to address the challenge of effectively integrating learner data from multiple related courses to enhance LLM-driven chatbots for teaching a subject?
    \item The infrequent validity of the assumption of independent and identically distributed samples in educational data raises concerns. Are prompt engineering algorithms in LLM-driven chatbots capable of handling noisy data effectively?
    \item Distinguishing between engagement and learning data will be crucial. The challenges of physical distancing in the online aspect of a flipped classroom can hinder effective communication between teachers and students. The absence of engagement pattern data may worsen inequity issues in learning, potentially disrupting the teaching and learning process with consequential effects \cite{mitpress}. LLM-based classroom flipping thus should enhance the communication between students and the instructor. How can LLM-driven chatbots learn classroom interaction pattern to mitigate these inequity issues and promote more inclusive learning experiences?%\footnote{Influenced by George Bernard Shaw's ``Pygmalion," Joseph Weizenbaum at MIT created Eliza, a chatbot, to explore human-machine communication, mirroring Professor Henry Higgins' attempt to refine Eliza Doolittle's speech in the play. Fast-forwarding over a century from Shaw's original work, the emergence of ChatGPT is poised to usher in a new era of chatbots capable of augmenting and elevating human skillsets.}
    \item Is it possible to incorporate LLM-driven chatbots into various aspects of education? While ChatGPT represents a significant advancement in NLP, it is crucial for AI technologies to steer clear of the ``successful first step" fallacy \cite{dreyfus}. Initial success in a controlled environment may not necessarily translate to realistic challenges. For instance, can LLM-driven chatbots be effectively utilized for automated assessment feedback, as explored in our preliminary study on automated peer grading \cite{autograding,Ling2}?
    \item How should domain-specific languages be integrated with LLM-driven chatbots for teaching? It is possible to extend the exploration using LLMs like GPT-3 and Codex to perform a specialized task, even for humanities subjects (e.g., to compose music or write a piece of journalism). Particularly, the use of ``LLM functions" to automate the interaction between the chatbot and LLMs using application programming interface (e.g., OpenAI ChatGPT API \cite{openai_2023}) will be essential to unlock the full potential of LLM-driven chatbots to enhance AI-assisted teaching.
    \item How should effective LLM-based classroom flipping techniques be designed to balance the tradeoffs between content capacity and comprehension capacity in teaching?
\end{itemize}

\section{Conclusion} 
Generative AI, like Large Language Models (LLMs), enables flexible and scalable pedagogical innovations such as flipped classrooms. This paper has introduced an LLM-driven flipped classroom method that combines Peer Instruction and Just-In-Time Teaching (JiTT) through reciprocal peer questioning. We have developed a workflow to incorporate prompt engineering in teaching and also an LLM-powered chatbot to integrate peer-generated questions and JiTT routines for student engagement and teaching adjustments for hybrid teaching, aiming for a unified evaluation of LLM's impact on teaching and learning, as opposed to a fragmented assessment.

For instructors embarking on new courses, this student-centric peer questioning approach serves as an efficient way to swiftly build a repository of clicker polls and quiz content. By tapping into collective student efforts to contribute to the question bank using LLMs, this approach not only lightens the instructor's workload but also encourages a well-rounded strategy that blends individual LLM-supported learning with interactive peer involvement during class. This cultivates students' ownership of their learning process and nurtures self-regulated learning skills, while aiding educators in addressing blind spots and allowing teachers to gain effective insights from teaching just in time, allowing them to adjust their teaching pace accordingly. 

%\section*{Acknowledgment}
%The author thanks Geoffrey M. Voelker, R. Srikant, Dah-Ming Chiu, Jacques Perche, Y.C. Tay for many useful discussions. He also thanks Alex Ling from Nautilus Software Technologies and Henry Herrington from Princeton University for the system implementation of the Nemobot AI chatbot.
%This work was supported in part by the University Grants Committee's Teaching Development and Language Enhancement Grant Project 6000710.

% acmart
% \bibliographystyle{ACM-Reference-Format}
\bibliographystyle{IEEEtran}
\bibliography{aiclassroomflipping}
\end{document}